# Energy-Environment evaluation and Forecast of a Novel Regenerative turboshaft engine combine cycle with DNN application

Mahdi Alibeigi[a], Mohammadreza Sabzehali[a]

**Abstract**

In this integrated study, a turboshaft engine was evaluated by adding inlet air cooling and regenerative cooling based on energy-environment analysis. First, impacts of flight-Mach number, flight altitude, the compression ratio of compressor-1 in the main cycle, the turbine inlet temperature of turbine-1 in the main cycle, temperature fraction of turbine-2, the compression ratio of the accessory cycle, and inlet air temperature variation in inlet air cooling system on some functional performance parameters of Regenerative turboshaft engine cycle equipped with inlet air cooling system such as power-specific fuel consumption, Power output, thermal efficiency, and mass flow rate of Nitride oxides (NOx) including NO and NO2 has been investigated via using hydrogen as fuel working. Consequently, based on the analysis, a model was developed to predict the energy-environment performance of the Regenerative turboshaft engine cycle equipped with a cooling air cooling system based on a deep neural network (DNN) with 2 hidden layers with 625 neurons for each hidden layer. The model proposed to predict the amount of thermal efficiency and the mass flow rate of nitride oxide (NOx) containing NO and NO2. The results demonstrated the accuracy of the integrated DNN model with the proper amount of the MSE, MAE, and RMSD cost function for both predicted outputs to validate both testing and training data. Also, $R$ and $R^2$ are noticeably calculated very close to 1 for both thermal Efficiency and NOx emission mass flow rate for both validations of thermal efficiency and NOx emission mass flow rate prediction values with its training and its testing data.

**NOMENCLATURE**

| | | | |
|---|---|---|---|
| $A$ | Area (m²) | **Greek symbol** | |
| $Cp$ | Specific heat capacity (J/kgK) | ΔT | Temperature variation (K) |
| $CR$ | accessory cycle compression ratio | $\eta$ | Thermal efficiency |
| $d$ | Density (kg/m³) | $\gamma$ | Speed of sound (m/s) |
| $f(x)$ | Activation function | $\phi$ | Equivalent ratio |
| $H$ | Flight Altitude (m) | **Subscript** | |
| $k$ | Specific heat ratio | a | Inlet air |
| $Ma$ | Flight-Mach number | c | compressor |
| $m$ | Mass flow rate (kg/s) | PC | Cooling Pump |
| $m$ | Mass flux (kg/s.m²) | $Comb$ | Combustor |
| $P$ | Pressure (kPa) | H | Heating |
| $r$ | Compression ratio of compressor | F | Fuel |
| R | Pearson Correlation factor | T | turbine |
| $R^2$ | Determination factor | **Abbreviation** | |
| $T$ | Temperature (K) | $MAE$ | Mean-absolute error |
| $Thrust$ | Thrust Force (kN) | $MSE$ | Mean-Squared Error |
| Q | Heat rate value | $RMSD$ | root-mean-squared deviation |
| TF | Temperature Fraction | $LHV$ | Lower heating value |
| $V$ | Velocity (m/s) | $PSFC$ | Power-Specific fuel consumption |
| $U_i$ | Actual output | GT | Gas Turbine |
| W | Power output (kW) | NOx | Nitride oxides |

1. **Introduction**

1.1. Background

   Gas turbine engines are one of the types of internal combustion engines that operate based on the Brayton cycle. Some aero engines are of the type of gas turbine (GT). Since the gas engine plays an important role in different industries such as electricity generation, and the air transmission industry. So, to improve the performance of gas turbine engines, optimization, prediction, and their thermodynamic cycle behavior are crucial in improving the performance of gas turbine engines. The most important performance parameters that are considered in thermodynamic cycle analysis can be output power, specific fuel consumption (PSFC), and thermal efficiency. The simplest procedure for an industrial gas turbine engine is the turboshaft engine [1, 2].

The turboshaft engine works like another gas turbine engine, which is named the turboprop engine, while the difference is that the hot gases are expanded into an open turbine or power turbine to produce more shaft power. Turboshaft engines are intended to produce shaft power only and are exploited in helicopters, ships, trains, tanks, pump units, numerous industrial gas turbines, and other implementations [3].

In this type of gas turbine engine, the airflow enters a compressor and enters the combustor so much that it reacts with fuel and then enters the turbine and makes a move in the turbine, then it exhausts to the ambient. The production power of the turbine is used to consume the power of the compressor and to turn on the generator (in aero engines, power shaft). It can be found that the prediction and evaluation are the cases based on future studies on such engines.

Machine Learning (ML) and Deep learning (DL) are the branches of Artificial intelligence (AI) that can be exploited in many fields associated with energy systems [4]. The gas turbine as one of the energy systems is no exception to this rule in order to use artificial intelligence [5]. The most significant, recurrent, and enhancing AI application in the gas turbine studies has been pointed out that anyone interested in this subject can find and find a better considerate. Contemplate using these models.

1.2. Literature Review

Recent studies verified that thermodynamical analysis is required to study gas turbine evaluation.

Aygun [6] attempted to analyze the thermodynamic cycle of an engine under different working conditions. In Aygun's study, the engine PSFC and output power are calculated to range between 377.5 to 588 grams per kilowatt hour and range between 377.33 and 1772.53 kW, respectively. Patel et al. [7] tried to optimize the subsonic /supersonic turbojet engine in the design point conditions based on the efficiency, the specific fuel consumption, and the thrust, the optimum value of the thermal efficiency and propulsive efficiency were respectively deliberated as 70.95 and 60.23 %,  Also, In another study, Zhou et al. [8] simultaneously examined and optimized recuperated gas turbine with a three-shaft engine and by decoupling power turbine mutable area nozzle (VAN) angle. They projected the control strategy for turbine shaft speed and VAN angle. When relative high-pressure shaft speed is respectively 0.95, 0.90, and 0.85, output power increased 6.37%, 15.88%, 47.80% and thermal efficiency increased 10.84%, 25.59%, 64.97% respectively.

Increasing the thermal efficiency of gas turbine engines has always been one of the challenges of the optimization of these systems. One of the conventional methods for increasing the efficiency of gas turbine engines is inlet air cooling. The results are shown by reducing the input air cooling temperature in gas turbine engines, thermal efficiency is increased. For instance, Najjar and Abubaker [9] demonstrated that by decreasing the inlet air temperature, inlet air density is increased so the inlet air mass flow rate is enlarged, which affects performance parameters such as thermal efficiency and thrust.

Baakeem et al. [10] inspected the development of the gas turbine power plant by inlet air cooling methods (TIAC) in optimum conditions with the cooling capacity of 36 kW/m$^3$s$^{-1}$ for all TIAC systems and an inlet temperature drop of 8 Celsius.  In another study, Yazdi et al. [11] investigated different climates in the inlet air of gas turbine engines in different climates of cities, including Yazd (hot–arid), Bandar Abbas (hot–humid), Ardabil (cold-humid), and Sari (humid subtropical). the cooling system methods consist of an absorption chiller, heat pump, and input mist eliminator system for inlet air cooling. In their study, the optimum working cooling method is based on different objective functions, such as pollution and the cost of electricity generation for each climate. Also, in another study, Deymi-Dashtebayaz and Kazemiani – Najafabad [12] explored the effect of different inlet air cooling methods of compressors in gas turbine engines. These methods include chiller media, mist eliminator, and absorption as normal cooling methods and pressure drop stations as a new cooling method. Part of the results of this study suggested that the absorption chiller system has the highest temperature drop in the air inlet temperature of the compressor, increasing the thermal and exergy efficiency of the cycle reported by about 2.5 and 3 %, respectively.

One of the common methods to increase GT thermal efficiency is adding heat recovery and combing it with the Rankine power generation cycle. These methods increase the thermal efficiency of the overall cycle by reducing the heat loss of the gas turbine cycle.
Bontempo and Manna [13] added an intercooler and a reheat Rankine cycle to the gas turbine engine and with regard to adding this advanced gas turbine cycle, they tried to optimize this specific advanced gas turbine based on power and efficiency. They determined that the energy efficiency of this advanced gas turbine is 24.96% higher than the simple one. Furthermore, in another study, Cha et al. [14] considered the effect of thermodynamic analysis of a GT engine cycle using inlet air cooling and heat recovery. The results showed that by adding the cooling in the inlet air and heat recovery, the output power and thermal efficiency increased to 25.4 % and 11.5% respectively. Sanaye et al. [15] analyzed, including Energy, Exergy, Economic, and Environmental (4E analysis), a plant that combined cooling, heating, power, and water system (CCHPW) involving a gas turbine, an Heat Recovery Steam Generator (HRSG) including steam and absorption refrigeration system equipped with an inlet air cooling system.
Recent studies have shown that hydrogen [16-18] used as fuel has advantages over the use of hydrocarbon fuels in gas turbine engines such as kerosene [19-21]. For instance, Derakhshandeh et al. [22] compared a hydrocarbon-fueled turbofan and a hydrogen-fueled turbofan based on the GE90 turbofan engine. Then an optimization with regards to economy and ecology was done. The consequences specified that in the hydrogen-fueled optimized cycle, the thermal efficiency improved by 2.65%. Kaya et al. [23] examined the advancement potential of the exergetic sustainability of a hydrogen-fueled turbofan UAV with heat recovery to save fuel and environmental influence reduction in the standard atmosphere with a relative humidity of 60% and altitude in the range of 0 to 16 km. In this way, fuel has been saved by about 11%.
Farahani et al. [24] investigated the thermal efficiency, exergy efficiency, and exergy destruction of the TF30-P414 turbofan engine to check the effect of the changes in the flight altitude (H) and the flight-Mach number (Ma) by using an Optimization approach.
The nitric oxide (NOx) formation Correlations are industrialized and clarified as a function of numerous parameters that affect performance and emissions [25].
Another challenge for a gas turbine engine is to reduce the mass flow rate of nitric oxide compounds emission in such systems. Therefore, a lot of efforts have been made in this regard. Park et al. [26] empirically investigated the reduction of NOx emission of a real gas turbine based on Mitsubishi Hitachi Power System's GT model for power generation: 501G by considering the pilot fuel split ratio of combustion-tuning parameters as sensitivity analysis for CCPP (combined cycle power plant) at Korea Western Power.

Zhang et al. [27] executed multi-objective genetic algorithm optimization with an objective function of higher efficiency and lower emissions for a helicopter turboshaft engine under given flight conditions with respect to the designated heat transfer surface geometries.

In recent years, artificial intelligence has been a hot topic for prediction for several applications, such as energy systems. For Example, Motaghian et al. [28] applied the Artificial Neural Network to predict the performance prediction of a solid desiccant wheel; also, it can be applied to predict in many fields such as prediction of Composite Properties [29, 30], prediction of Tensile − Shear strength [31], and prediction of temperature and force in bone drilling process [32]. However, the most common usage of ANN in the gas turbine is fault diagnosis. For instance, Bai et al. [33] investigated the use of a Convolutional neural network (CNN) based on deep transfer learning to detect the fault detection of data-rich gas turbine combustion chambers.

The other application of artificial intelligence in the gas turbine is the prediction of performance or pollutant emissions. For instance, the prediction of turbofan engine performance as one of the gas turbine engines has been studied by Sabzehali et al. [34]; in another study, Liu and Karimi [35] predicted a GT performance via machine learning. The result demonstrated that the average and maximum errors were calculated at less than 2.0% and 4.3%, respectively. Pierezan et al. [36] combined Artificial Neural Networks (ANN) and Cultural Coyote Optimization Algorithm (CCOA) for heavy-duty gas turbine performance. As a consequence, the CCOA developed the basic GT performance meaningfully, and specific consumption (TSFC) was reduced up to 3.6%. Kayaalp and Metlek [37] predicted successfully the pollutants such as CO, CO2, UHC, and NO2 that the absolute mean error values for each one were 0.1473, 0.0442, CO2, 0.0369, and 0.0028, respectively.

The other usage of ANN can be found in studies of predicted NOx emissions; in this regard, Wang et al. [38] predicted NOx emissions and NO mass flow rate MFR with various input selections by means of mutual information (MI) and back propagation neural network (BPNN). In their methodology based on the Ml-feature selection, the BPNN mean absolute deviation (MAD) and the BPNN root mean square error (RMSE) were decreased by almost 15%, which also indicated that the MI-feature selection method was effective. The investigation demonstrated that it can be effective, lower computational cost, and be useful to control the after-treatment system of real driving vehicles.

Dirik [39] used a hybrid intelligence system to predict NOx emissions. For this purpose, the Adaptive Neuro-Fuzzy Inference System (ANFIS) model was optimized by utilizing a genetic algorithm (GA) to decrease the errors such as mean squared error (MSE), the error means (EM), the root-mean-squared error (RMSE), the standard deviation (STD), and mean absolute percentage error (MAPE). The results showed that the minimum values of error parameters such as MSE, RMSE, EM, STD, and MAPE were calculated to be 24.8379, 4.9838, 3.4625e-05, 4.9839, and 5.1660, respectively, for the training data. The minimum values of these error parameters for the test data were found as 26.5961, 5.1571, 0.065696, 5.157, and 5.3695 for ANFIS-GA, respectively.

### 1.3. Contribution and organization

As the recent literature reviewed, even though performance prediction [34, 35] and NOx emission [38] were studied separately for engines, it needs to be used in the new concept of engines such as turboshaft engines as one of aero gas turbine engines to predict both its NOx emission and its performance that never has done. First, the prediction needs the calculated data based on the Energy-Environment analysis of the new concept turboshaft engine. In this integrated study, in order to increase the thermal efficiency and output power, and also reduce the production of nitrogen oxides (NOx) produced by the turboshaft engine, adding an inlet air cooling [14, 40] system and a regenerative [41-43] cooling system and also considering liquid hydrogen as a fuel has done thermodynamically and environmentally for a new concept turboshaft engine that was never exploited before. Then a deep neural network is modeled to predict the NOx emission mass flow rate and thermal efficiency at the same time as to set the predicted data to the real data gained from the energy-environment analysis of the new Regenerative turboshaft engine combined cycle. Eventually, current study could be opened the novel intellect turboshaft engine with deep neural network processing and equip the inlet air cooling and regenerative system at the same time.

## 2. Methodology

In this study, the case study consists of two combined cycles together; the cycle of an open gas turbine cycle as a turboshaft engine can be called the main cycle. In this cycle, the input airflow first begins by passing through the inlet air exchanger (Heat exchanger 1), so airflow density increases, and then enters the compressor. The airflow compresses in the compressor 1 (c1) and is ready to enter the combustion chamber; after reacting with fuel, the temperature and enthalpy increase to reach the inlet temperature of turbine 1 (TIT), so it enters turbine 1, and after rotating the turbine 1 shaft, it enters to heat recovery of exhaust flow (Heat exchanger 2); after reduction of temperature. It exhausts the ambient.

The other cycle is the sub-cycle flow called the accessory cycle. A close Rankine cycle is associated with liquid nitrogen as a working fluid. In this case, the liquid nitrogen enters heat exchanger 1 after increasing the pressure at the cooling pump (PC) and cools the input airflow, and the sub-cycle flow enters turbine 2 and reduces the temperature of the heat exchanger. After cooling the exhaust flow, it reaches the temperature of turbine 3, then it rotates to the shaft turbine 3. So, it enters heat exchanger 3 (hex3). In heat exchanger 3, using liquid hydrogen fuel flow as a working refrigerant fluid for the accessory cycle, caused by decreasing temperature to the boiling temperature of the nitrogen in the turbine 3 output pressure; liquidity transpires for nitrogen. Then liquid nitrogen flow enters the PC pump again.

The thermodynamics (energy analysis) and environmental results were obtained from Engineering Equation Solver (EES).

The diagram of the novel regenerative turboshaft engine combined cycle is illustrated in Fig. 1.

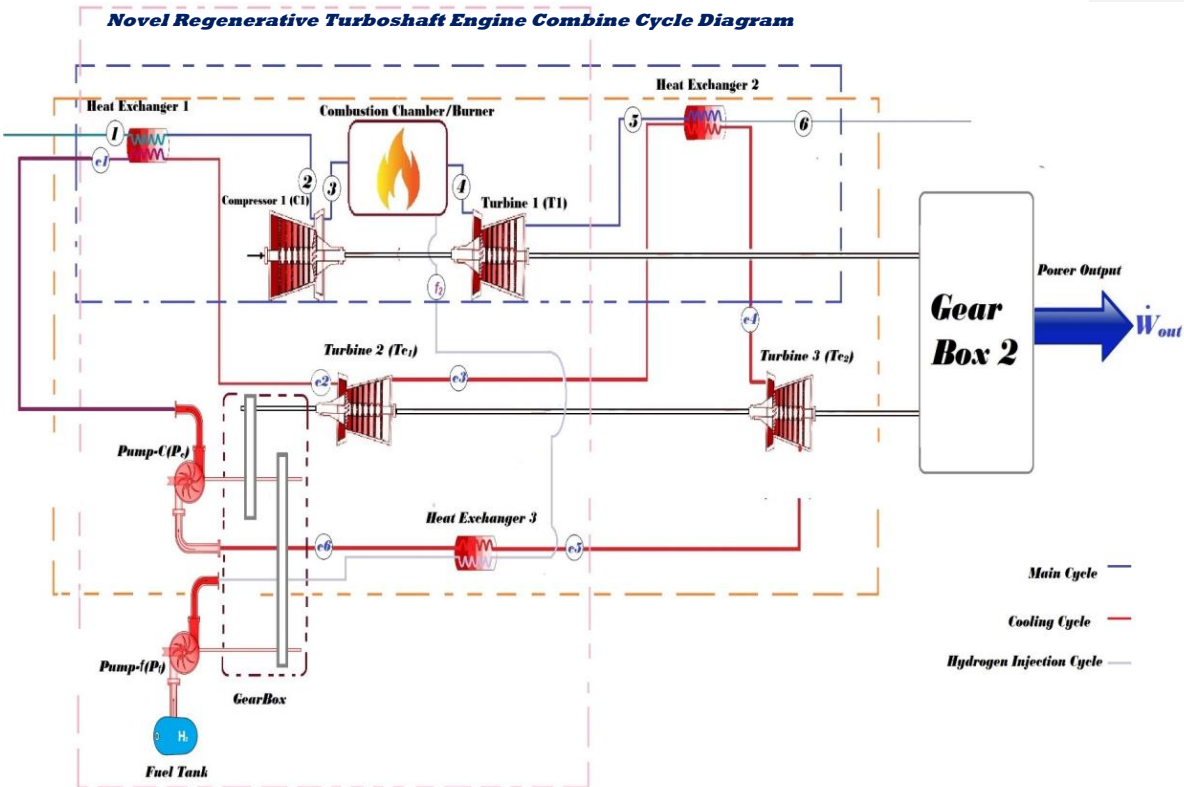

**Fig. 1.** Novel Regenerative turboshaft engine combines cycle Visualization

2.1. Thermodynamic modeling of the main cycle (Gas turbine cycle)

In this case, a turboshaft gas turbine cycle as the main cycle combined with a Rankine cycle in order to improve the inlet air and heat recovery of exhaust flow as a sub-cycle was studied. Pressure drops of airflow in the inlet air cooling and upstream heat exchanger 1 (hex1) flow are negligible, it can be approved for pressure in eq (1);

$$P_2 = P_1 \quad (1)$$

$$T_2 = T_1 - \Delta T_{cool} \quad (2)$$

where $P_1$, $T_1$, $P_2$, and $T_2$ are the inlet pressure of heat exchanger 1, inlet temperature of the heat exchanger 1, and the pressure and temperature of the inlet air compressor, respectively; $P_3$ and $T_3$ are the pressure and temperature of the inlet air at the burner. $\Delta T_{cool}$ is the temperature variation between the inlet air temperatures that varies according to the temperature of the environment.

The c1 outlet pressure ($P_3$) is calculated as:

$$P_3 = r_{c1} P_2 \quad (3)$$

where $r_{c1}$ is pressure ratio of compressor and $P_2$ is the pressure of the inlet air compressor.

The outlet temperature of the compressor ($T_3$) is obtainable as [44];

$$T_3 = T_2 \left[ 1 + \frac{1}{\eta_{c1}} \left( r_{c1}^{\frac{k_{c1}-1}{k_{c1}}} - 1 \right) \right] \quad (4)$$

where $\eta_{c1}$ and $k_c$ are isentropic compressor efficiency and the friction of special heat capacity in the pressure constant process to the special heat capacity in the volume constant process in the compressor, which is the function of the mean temperature of the passing air from along the compressor, also the compressor consumed power ($W_{c_1}$) can be premeditated as;

$$W_{c1} = m_a T_2 \left[1 + \frac{1}{\eta_{c1}} \left(r_{c1}^{\frac{k_{c1}-1}{k_{c1}}} - 1\right)\right] \quad (5)$$

where the inlet air mass flow rate to the engine ($m_a$) measured as;

$$m_a = d_a V_a A_a \quad (6)$$

where $A_a$, $d_a$, and $V_a$ are the surface of the air inlet, air density, and inlet airflow velocity, respectively.

Fuel heat value rate ($Q_h$) is intended as;

$$Q_h = m_a C_p (T_4 - T_3) \quad (7)$$

Where $C_p$ is the heat capacity of air at constant pressure as a function of temperature, which is defined as [45];

$$Cp = a_1 + a_2 T^1 + a_3 T^2 + a_4 T^3 \quad (8)$$

Air is combined with a molar composition of 21% oxygen and 79 % 183 nitrogen and other gases. Also, $a_1$, $a_2$, $a_3$, and $a_4$ are constant values.

In this case, constants $a$ values must be used as;

$$\begin{cases} a_1 = 0.99963438 \\ a_2 = -0.055205312 \times 10^{-3} \\ a_3 = 0.346320281 \times 10^{-6} \\ a_4 = -0.140118997 \times 10^{-9} \end{cases} \quad (9)$$

Also, $T_4$ is the inlet flow temperature in the turbine (1) and $T_3$ is the outlet flow temperature in the compressor.

Correspondingly, the fuel consumption mass flow rate can be determined as [46];

$$m_f = \frac{Q_h}{LHV \, \eta_{comb}} \quad (10)$$

where $LHV$, $\eta_{comb}$, and $Q_h$ are the lower heating value per kilogram of fuel, combustion efficiency, and heat value rate, respectively.

The summation of these two-mass flow rates defined the passing mass flow rate through the turbine as;

$$m_T = m_a + m_f$$
$$m_T = d_a V_a A_a + \frac{Q_h}{LHV\,\eta_{comb}} \tag{11}$$

The generated power of turbine 1 can be defined as the following equation.

$$W_{T1} = m_{T1}.Cp_{T1}.(T_4 - T_5) \tag{12}$$

Where $m_{T1}$ and $Cp_{T1}$ are the mass flow rate of passing from turbine 1 and specific heat capacity at pressure constant.

According to the principle of conservation of energy, it is supposed that mechanical power loss in shafts is ignored.

Also, the heat traded in the heat exchanger 2 is calculated as;

$$Q_2 = m_a Cp_5 (T_5 - T_6) \tag{13}$$

where $T_5$ and $T_6$ are the input temperature and the output of the heat exchanger 2; also, $Cp_5$ is specific heat compacity at constant pressure in the temperature $T_5$ and the pressure $P_5$.

Also, the net power of the main cycle is calculated.

$$W_{net1} = W_{T1} - W_{c1} \tag{14}$$

where $W_{T1}$ and $W_{c1}$ are the generation power of turbine 1 and compressor consumed power of the main cycle, respectively.

Flight Mach number (F-Mach) can be obtained as;

$$Ma = \frac{V}{\gamma} \tag{15}$$

where $V$ and $\gamma$ are local flow velocity and sound speed, respectively.

2.2. Thermodynamic modeling of the accessory cycle (inlet air refrigeration system)

In this case, according to the cycle schematic in Fig. 1; a subsystem was added to improve the inlet air and recover the exhaust flow heat from a Rankine cycle with liquid nitrogen as a working fluid. Also, the sub-system cycle is called the accessory cycle.

The heat converted in the heat exchanger 1 is calculated as;

$$Q_1 = m_a Cp \Delta T \tag{16}$$

where $m_a$, $Cp$ and $\Delta T$ are the mass flow rate of inlet air in the refrigeration cycle, the specific heat rate capacity in the pressure constant approach and the different temperatures of the inlet air refrigeration cycle in the engine to the ambient temperature.

The accessory cycle compression ratio ($CR$) can be considered as;

$$`CR = \frac{P_{c1}}{P_{c6}} \tag{17}$$

where $P_{c6}$ and $P_{c1}$ are the inlet and outlet pressure of the pc pump, respectively.

The inlet power ($W_{PC}$) to the Pc pump is considered as;

$$W_{PC} = \frac{m_{PC} P_{c6}(CR-1)}{\eta_{PC} d_c} \tag{18}$$

where $P_{c6}$, $\eta_{pc}$, $d_c$, and $m_{PC}$ are the inlet pressure of the pc pump, pc-pump isentropic efficiency, refrigerant fluid density, and refrigerant fluid mass flow rate, respectively.

The output temperature of the heat exchanger 1 ($T_{c2}$) is calculated i.e.;

$$T_{c2} = T_{c1} + \left(\frac{Q_1}{m_c Cp_{c2}}\right) \tag{19}$$

where $T_{c1}$ and $Cp_{c2}$ are input temperature of heat exchanger 1 and heat capacity of refrigerant fluid at constant pressure, respectively.

Heat exchanger 1 exit pressure is obtained as follows;

$$P_{C2} = P_{C1} - \Delta P_{ch1} \tag{20}$$

where $P_{C1}$, $P_{C2}$, and $\Delta P_{ch1}$ are inlet downstream heat exchanger 1 pressure, exit downstream heat exchanger 1 pressure, and pressure drop in the Heat exchanger 1, respectively. $\Delta P_{ch1}$ evaluated as a percentage of the air exit pressure of the compressor.

Also, the output temperature of turbine 2 ($T_{c3}$) is intended as;

$$T_{c3} = T_{c2}TF_{T2} \tag{21}$$

where $TF_{T2}$ and $T_{c2}$ temperature fraction of turbine 2 and the inlet temperature of turbine 2, respectively; also, the outlet pressure of turbine 2 ($P_{c3}$) can be calculated as;

$$P_{c3} = P_{c2}\left((TF_{T2})^{\frac{K_{C2}}{K_{C2}-1}}\right) \tag{22}$$

where $P_{c2}$, $K_{C2}$, $T_{c2}$, and $TF_{T2}$ are the inlet pressure of turbine 2 and the ratio of the special heat capacity, which is the division of specific heat capacity at constant pressure to the specific heat capacity at the constant volume, inlet the temperature of turbine 2, and temperature ratio of turbine 2, respectively.

In heat exchanger 2, the exit temperature of the heat exchanger is defined as;

$$T_{c4} = T_{c3} + \left(\frac{Q_2}{m_c Cp_{c3}}\right) \tag{23}$$

Where $T_{c3}$, $T_{c4}$, $Cp_{c3}$, and $T_{c3}$ are inlet temperature of the heat exchanger 2, output temperature of the heat exchanger 2, and refrigerant fluid heat capacity at constant pressure in $T_{c3}$ temperature and $P_{c3}$ pressure.

Heat exchanger2 exit pressure is attained as follows;

$$P_{C4} = P_{C3} - \Delta P_{ch2} \tag{24}$$

where $P_{C3}$, $P_{C4}$, and $\Delta P_{ch2}$ inlet heat exchanger-1 pressure, exit Heat exchanger-2 pressure, and pressure drop in the Heat exchanger 2, respectively; also, $\Delta P_{ch2}$ is evaluated as a percentage of the air exit pressure of the compressor.

Also, exit turbine 3 pressure ($P_{c5}$) is calculated as;

$$P_{c5} = P_{c4}\left(\left(\frac{T_{c5}}{T_{c4}}\right)^{\frac{k_{c4}}{k_{c4}-1}}\right) \tag{25}$$

where $P_{c4}$ and $k_{c4}$ are the inlet pressure of turbine 3 and the ratio of special heat capacity at constant pressure to the special heat capacity at the constant volume in temperature and pressure $T_{c4}$ and $P_{c4}$, respectively; also, $T_{C5}$ is the output temperature of the turbine 3.

The exchanged heat value ($Q_3$) is calculated as;

$$Q_3 = m_c Cp(T_{C5} - T_{C6}) \tag{26}$$

Where $m_c$ is the mass flow rate of the cooling fluid, $Cp$ is specific heat rate capacity in pressure constant approach; also, $T_{C6}$ is the temperature of the cooling fluid that is equal to the boiling temperature of the cooling fluid in $P_{C6}$ pressure.

exit heat exchanger 3 pressure is obtainable as follows;

$$P_{C6} = P_{C5} - \Delta P_{ch3} \tag{27}$$

Where $P_{C5}$, $P_{C6}$, and $\Delta P_{ch3}$ are inlet heat exchanger 1 pressure, exit heat exchanger 3 pressure, and pressure drop in the Heat exchanger 3, respectively; also, $\Delta P_{ch3}$ is evaluated as a percentage of the air exit pressure of the compressor.

The inlet power to the fuel pump (PF pump) is defined as;

$$`W_{Pf} = \frac{m_f(P_{f2} - P_{f1})}{\eta_{pf} d_f} \tag{28}$$

where $W_{Pf}$, $P_{f1}$, and $P_{f2}$ the power consumption of the pf pump, inlet pressure of pf pump, output pressure of pf pump, respectively; also, $\eta_{pf}$, $d_f$, and $m_f$ are isentropic efficiency of pf pump, refrigerant fluid density, and fuel consumption mass flow rate, respectively.

Also, the output power of turbine 2 is calculated as;

$$W_{T2} = m_C C_{PC2}(T_{C3} - T_{C2}) \tag{29}$$

where $C_{pC2}$ is refrigerant fluid-specific heat capacity in $T_{c3}$ temperature and $P_{c3}$ pressure.

It is assumed that mechanical power loss in shafts is negatable of loss according to the principle of conservation of energy. So, the outlet power of turbine 3 is intended as;

$$W_{T3} = m_C C_{PC4}(T_{C4} - T_{C5}) \tag{30}$$

where $C_{pC4}$ is the specific heat capacity of refrigerant fluid in $T_{c4}$ temperature and $P_{c4}$ pressure.

Notably, it is assumed that the ignorance of loss of mechanical power in shafts is due to the principle of conservation of energy.

Also, the accessory output power is defined as;

$$W_{net2} = W_{T2} + W_{T3} - W_{PC} - W_{Pf} \tag{31}$$

where $W_{Pf}$, $W_{PC}$, $W_{T3}$, and $W_{T2}$ are the power consumption of the fuel pump and the power consumption of the pump in the accessory, the power generation of turbines 3 and 2, respectively.

The output power ($P_{out}$) is equal to the sum of the power of the main and accessory cycles, which is calculated as;

$$P_{out} = W_{net2} + W_{net1} \tag{32}$$

where $W_{net1}$ and $W_{net2}$ are the net generating power of the main and accessory cycle, respectively.

Power-specific fuel consumption is known as the specific fuel consumption of an engine, which is defined as the friction of mass flow rate of fuel consumption ($m_f$) to output power of engine in the following equation, PSFC can be calculated as;

$$PSFC = \frac{m_f}{P_{out}} \tag{33}$$

$\eta_{th}$ is the thermal efficiency of the engine which is calculated as;

$$\eta_{th} = \frac{P_{out}}{Q_h} \tag{34}$$

where $Q_h$ is the heat rate value.

The mass flux for each species is defined as;

$$\bar{\bar{m}}_i = \frac{\dot{m}_i}{A} \tag{35}$$

where $\dot{m}_i$ and $A$ are the mass flow rate of species i and area, respectively.

Consider a non-reactive gas mixture consisting of only two molecular species, species A and B. Fick's law is defined as spreading through others by one-dimensional binary diffusion ($Y_A + Y_B = 1$), where $Y_A$ and $Y_B$ are the mass fraction of species A and species B, respectively. On the other hand, the diffusional flux of species A and B should be according to the following equation for binary mixtures at one dimension.

$$-\rho D_{AB}\frac{dY_A}{dx} - \rho D_{BA}\frac{dY_B}{dx} = 0 \tag{36}$$

where $\rho$, $D_{AB}$, and $D_{BA}$ are the gas mixture density, and diffusion coefficient of A diffused through B and B diffused through A, respectively.

Individually, the mass flux for each species can be defined as;

$$\bar{\bar{m}}_i = Y_i\bar{\bar{m}} - \rho D_i\frac{dY_i}{dx} \tag{37}$$

So, the mass flow rate for each species can be calculated by the mass fraction gotten as;

$$\dot{m}_i = y_i m_{tot} \tag{38}$$

Besides, the $NO_x$ emission has two Parts: $NO$ and $NO_2$. Accordingly, the mass flow rate of each emission can be determined as follows;

$$\dot{m}_{NO} = y_{NO} m_{tot} \tag{39}$$

$$\dot{m}_{NO_2} = y_{NO_2} m_{tot} \tag{40}$$

Consequently, the $NOx$ emission mass flow rate can be determined as;

$$\dot{m}_{NO_x} = \dot{m}_{NO} + \dot{m}_{NO_2} \tag{41}$$

2.3. $NO_x$ formation

The hydrogen fuel has no nitrogen, but in the combustion products of the combustion chamber, the nitride oxide compound ($NO_x$) is emitted to the ambient.

The term $NO_x$ is a chemical compound for molecules comprehending at least one oxygen atom with one nitrogen. $NO_x$ is composed of nitrogen gas in the air in all ambient. Most of $NO_x$ generation is nitrogen oxide ($NO$), with a miniature proportion of nitrogen dioxide ($NO_2$), and a slight measure of other $NO_x$.

$NO_x$ emission from the exhaust appearances reacts in the environment to form an ozone layer. This is one of the crucial motives for studying the formation of $NO_x$ [47].

Nitrogen can similarly be found in fuel mixtures, which may comprise minor measures of ammonia $NH_3$, cyanide $NC$, and Hiroden cyanide $HCN$. There are numerous imaginable reactions that surround $NO$, which are all probable to transpire through the combustion process in an immediate duration [48].

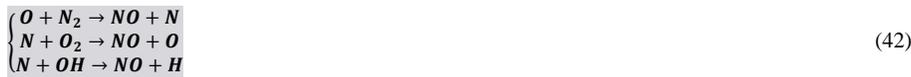
$$\begin{cases} O + N_2 \rightarrow NO + N \\ N + O_2 \rightarrow NO + O \\ N + OH \rightarrow NO + H \end{cases} \tag{42}$$

$NO$ can supplementarily react to the shape form of $NO_2$.

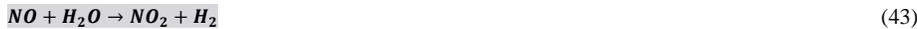
$$NO + H_2O \rightarrow NO_2 + H_2 \tag{43}$$
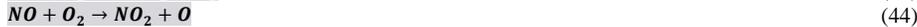
$$NO + O_2 \rightarrow NO_2 + O \tag{44}$$

$NOx$ is composed of four mechanisms that contain nitrogen from the air [49]:

a. Thermal mechanism (Zeldovich)
b. Prompt mechanism (Fennimore)
c. $N_2O$ intermediate mechanism
d. $NNH$ mechanism (fuel $NO$)

### 2.3.1. Zeldovich mechanism

$NO$ is formed at the high temperature above the range of 1600–1700 K. Above this range, despite these temperatures, $NOx$ is formed in minute amounts. Also, $NOx$ starts to form inside the combustion chamber [50].

Originally, the rate of formation of $NO$ was defined by the use of the three reactions based on the Arrhenius equation.

**Table 1.** Arrhenius parameters for the $NO$ reaction based on the Zeldovich mechanics [51]

| | Reaction rate constant ($m^3/kg\ mol - s$) | |
|---|---|---|
| | Forward reaction | Backward reaction |
| $O + N_2 \leftrightarrow NO + N$ | $k_1 = 1.8 \times 10^{14} \times \exp(-\frac{38,370}{T})$ | $k_{-1} = 1.8 \times 10^{14} \times \exp(-\frac{425}{T})$ |
| $N + O_2 \leftrightarrow NO + N$ | $k_2 = 1.8 \times 10^{14} \times \exp(\frac{4680}{T})$ | $k_{-2} = 1.8 \times 10^{14} \times \exp(-\frac{20,820}{T})$ |
| $N + OH \leftrightarrow NO + H$ | $k_3 = 7.1 \times 10^{13} \times \exp(-\frac{450}{T})$ | $k_{-3} = 1.7 \times 10^{14} \times \exp(-\frac{24,560}{T})$ |

The $NO$ formation rate can be defined by using three reactions as;

$$\frac{d[NO]}{dt} = k_1[O][N_2] - k_{-1}[NO][N] - k_2[N][O_2] - k_{-2}[NO][O] + k_3[N][OH] - k_{-3}[NO][H] \qquad (45)$$

### 2.3.2. Fennimore mechanism

The Fennimore mechanism is confidentially accompanying the combustion chemistry of hydrocarbons ($HC$). $NO$ shape probably quickly in the flame reaction region.

The prompt $NO$ is formed in the flame by the response of provisional material types of $CN$ set with $O$ and $OH$ radicals. The $HC$ radicals $CH$, $CH_2$, $C$, $C_2$, etc. formed in the flame front rejoin with sub-atomic nitrogen to attain provisional species, for instance, $HCN$ and $CN$ by the reactions. Enormous concentrations of cyanides ($CN^-$) supplementary closer to the reaction flame have been inspected during a higher fuel-air (FA) equivalence ratio and quick $NO$ formation has been seen related to quick degeneration of $HCN$ [49].

$$CH + N_2 \leftrightarrow HCN + N$$
$$C + N_2 \leftrightarrow CN + N \qquad (46)$$

The hydrogen cyanide ($HCN$) converted into nitric oxide ($NO$) for $\Phi < 1.2$. The prompt $NO$ by considering the Fennimore mechanism is discovered to be closer to 10%. The combustion period in the engines is lower due to the higher combustion pressure [47].

The conversion sequence is obtainable as;

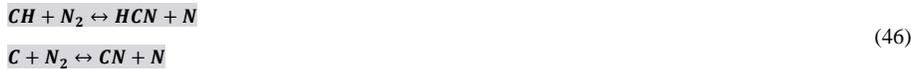

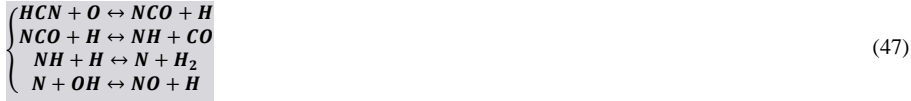

(47)

### 2.3.3. $N_2O$ intermediate mechanism

$N_2O$ intermediate mechanism is typically used by gas turbine industries in engines operating under fuel-lean ($\Phi < 0.8$), and low-temperature conditions.

In gas turbine engines, it plays a crucial role in monitoring the $NO$ formation when the engine is operating under premixed combustion [49].

Notably, $\Phi$ is the equivalent ratio. The three steps of $N_2O$ intermediate mechanism is;

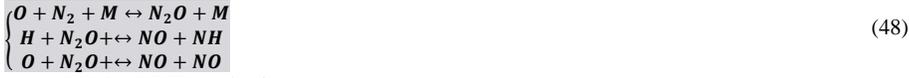

(48)

### 2.3.4. NNH mechanism

$NO$ formation based on the $NNH$ mechanism is the latest discovery of $NO$ formation in the reaction of nitrogen oxides. The nitrogen reacts to form at first with radical $H$ and then it is reactive with radicals. For example, $HCN$, $NH3$, $CN$, $NH$, etc. These species are further oxidized to $NO$.
The two key phases' reactions in this mechanism are contained as [47];

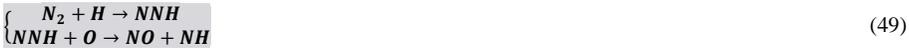

(49)

## 2.4. Deep Neural Network Approach

In recent years, deep learning techniques have provided powerful techniques for improving modeling and predictive performance [52]. Deep learning techniques use a deep or layered architecture. This is the basic structure of deep neural networks, and the main difference between deep neural networks (DNN) is the number of layers. In general, flat neural networks have only a few layers of neural networks, which limits their ability to express complex functions [53]. Conversely, deep learning has more than five neural network layers, providing a more efficient algorithm that can be even more accurate. The deep neural network method is regarded as a good ML method because it adds some hidden layers to a normal ML neural network, and is popular in various fields such as energy consumption prediction [54] and turbofan performance prediction [34]. However, there is a rare investigation that studied deep learning prediction of aero engines for both emission and performance.
In this study, the DNN was designed to predict the energy and environment parameters such as Thermal Efficiency and NOx emission mass flow rate. In this regard, the network considers 4 layers, including an input layer, 2 hidden layers, and an output layer. The parameters of the input layer are Turbine Temperature Fraction, Turbine Pressure fraction, Turbine Inlet Temperature, Pressure ratio, Difference Temperature, flight Mach number, and flight altitude (H). Also, the output layer is defined as Thermal Efficiency and NOx Emission mass flow rate. Each hidden layer has manually 625 neurons. Also, the loss function and optimizer are defined as the mean squared error and ADAM based on Keras library which included the TensorFlow library in the Python open-source scripting coding system. Consequently, the network has 397,502 parameters gained by Neurons and biases. The network is illustrated in Fig. 2.

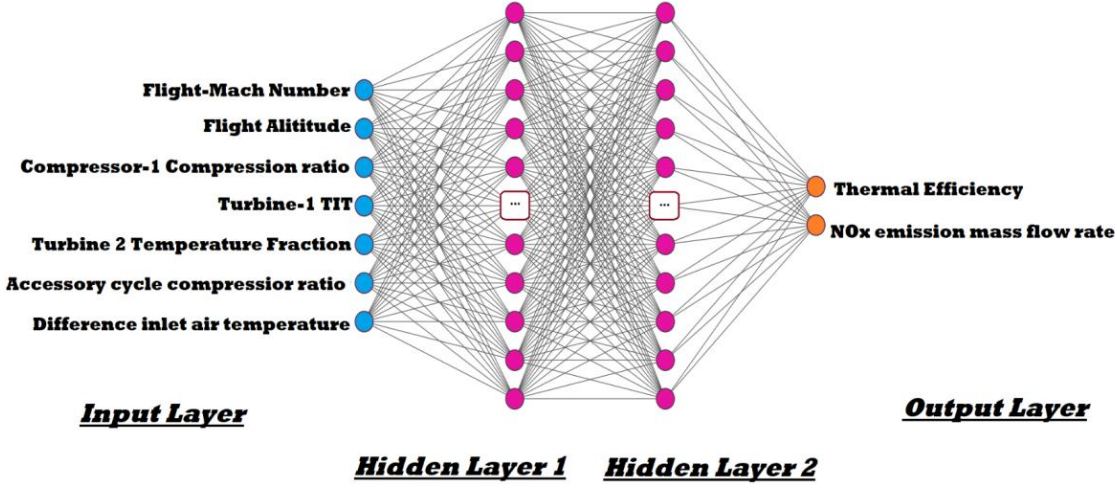

**Fig. 2.** Deep Neural Network Visualization

Some parameters are deliberated for comparison to achieve an accurate model. These parameters are described in the following equation.

The mean absolute error (MAE) can be defined as;

$$MAE = \frac{1}{n}\sum_{i=1}^{n}|U_i - \widehat{U}_i| \qquad (50)$$

The mean squared error (MSE) can be calculated as;

$$MSE = \frac{1}{n}\sum_{i=1}^{n}(U_i - \widehat{U}_i)^2 \qquad (51)$$

Root-mean-square deviation (RMSD) can be determined as;

$$RMSD = \sqrt{\frac{1}{n}\sum_{i=1}^{n}(U_i - \widehat{U}_i)^2} \qquad (52)$$

Pearson correlation factor $(R)$ and determination factor $(R^2)$ signify the correlation between the true values, especially for the test, and the predicted values of the DNN model. In this regard, if $R$ and $R^2$ are calculated closer to one, the predicted values are more matched to the true values.

$$R = \frac{\sum_{i=1}^{n}[(U_i - U_i^{mean})(\widehat{U}_i - \widehat{U}_i^{mean})]}{\sqrt{[\sum_{i=1}^{n}(U_i - U_i^{mean})^2][\sum_{i=1}^{n}(\widehat{U}_i - \widehat{U}_i^{mean})^2]}} \qquad (53)$$

$$R^2 = \frac{\left[\sum_{i=1}^{n}(U_i - U_i^{mean})(\widehat{U}_i - \widehat{U}_i^{mean})\right]^2}{\left[\sum_{i=1}^{n}(U_i - U_i^{mean})\right]\left[\sum_{i=1}^{n}(\widehat{U}_i - \widehat{U}_i^{mean})\right]} \tag{54}$$

where $U_i^{mean}$, $\widehat{U}_i^{mean}$, $U_i$, and $\widehat{U}_i$ are the mean values of true data (test data), the mean value of predicted data from the DNN model, the values of the i-sample for true data (test data), and the i-sample predicted value, respectively.

3. **Result and discussion**

The output power variations at a flight altitude are shown in Fig. 3a at constant F-Mach=0.5. The results demonstrate that the output power is reduced by increasing the flight altitude ranging from about 3000 meters to 4000 meters because, by reducing the air density due to increasing the flight altitude, the inlet air mass flow rate to the engine decreases and reducing the density of inlet air leads to decreasing the mass flow rate of the main cycle power turbine and consequently the reduction of the generated power of the main cycle power turbine. Also, the variation of the output power with the Flight-Mach number at a constant altitude of 4000 m is shown in Fig. 3b. Since the inlet air mass flow rate to the engine increases with an increase in the Flight-Mach number, accordingly, the output power increases. The variation of PSFC and the thermal efficiency of the cycle at a flight altitude at constant F-Mach=0.5 were shown in Fig. 3b.

The results confirm that with increasing the flight altitude, in the flight altitude ranges from about 3000 meters to 4000 meters at the constant Flight-Mach number, the thermal efficiency is increased and power-specific fuel consumption is reduced. Also, the variation of the power-specific fuel consumption and thermal efficiency at a constant flight altitude of 4000 m is shown in Fig. 3b. The results confirm that the constant flight altitude with a higher Flight Mach number in the flight Mach numbers ranges from about 0.3 to about 0.8 is caused an increase in thermal efficiency but it decreases PSFC.

Also, in Fig. 3c, the generated mass flow of $NO$ and $NO_2$ with varied flight altitudes at constant F-Mach= 0.5 is displayed. Results show that with a decrease in the generated mass flow rate of $NO$ and a decrease in $NO_2$ with Flight altitude increasing.

Subsequently, the generated mass flow rate of $NO$ and $NO_2$ with varied Flight-Mach numbers at a constant altitude of 4000 m is shown in Fig. 3c. The results show that with an increase, the Flight-Mach number on $NO$ and $NO_2$ production has increased.

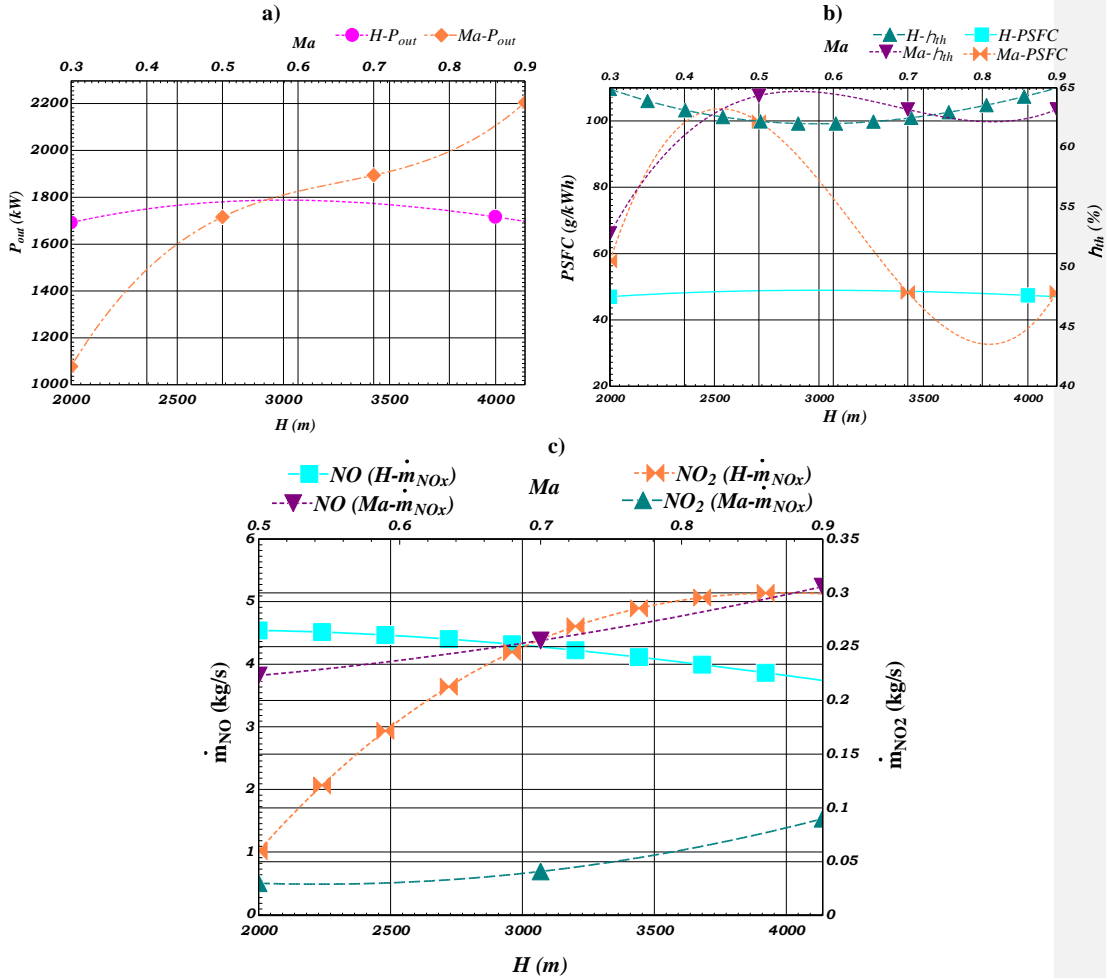

**Fig. 3.** Effect of Flight Altitude and Flight Mach Number on a) Power output b) PSFC and Thermal Efficiency c) NOx emission mass flow rate

In this section, the effect of parameters such as the main compressor compression ratio ($r_{c1}$), accessory cycle compression ratio ($CR$), inlet temperature of the main cycle power turbine ($TIT$), and turbine 2 temperature fraction ($TF_{T2}$), and the different temperatures between ambient air temperature and inlet compressor temperature ($\Delta T_{cool}$) on cycle performance parameters including the output power, PSFC, and thermal efficiency are investigated.

The output power variations for $TIT$ and the main cycle compression ratio at a flight altitude of 4000 m and F-Mach=0.3 are represented in Fig. 4a. The results show that the output power is increased due to an increase in $TIT$. Also, the results show that by increasing the compression ratio of the main cycle compressor, the output power cycle is increased.

The variation of power-specific fuel consumption and thermal efficiency with $TIT$ and the compression ratio of the main cycle turbine at a flight altitude of 4000 m and F-Mach=0.3 are shown in Fig. 4b. The results confirm that increasing $TIT$, power-specific fuel consumption increases but thermal efficiency decreases.

Also, the results show that by increasing the compression ratio of the main cycle compressor, the thermal efficiency is increased and specific fuel consumption is reduced.

Also, changes in $NO$ and $NO_2$ produced mass flow rate varied by $TIT$ and the compression ratio of the main cycle compressor indicated in Fig. 4c. Results showed an increase in the compression ratio of the main cycle compressor was caused by an increase in produced mass flow rate of $NO$ and $NO_2$. Similarly, $NO$ and $NO_2$ production mass flow rates increase with an increase in $TIT$.

Also, the output power variations for the inlet air temperature are shown in Fig. 5b. Results indicate that the output power is increased due to reducing the air temperature in the range $0\ K \leq \Delta T_{cool} \leq -25\ K$ and the range $-100\ K \leq \Delta T_{cool} \leq -50\ K$.

Also, the variation of power-specific fuel consumption and thermal efficiency with respect to the different temperatures is shown in Fig. 5a.

The results showed that in the range of $-100\ K \leq \Delta T_{cool} \leq -70\ K$, with inlet air cooling thermal efficiency increases, and the power-specific fuel consumption decreases. Also, with inlet air cooling in the range of $-70\ K \leq \Delta T_{cool} \leq -10\ K$, the thermal efficiency decreases, and $PSFC$ increases.

As well as changes in $NO$ and $NO_2$ produced mass flow rate and power output at different temperatures indicated in Fig. 5b. The results showed that with decreasing temperature variation ranging from $-100\ K \leq \Delta T_{cool} \leq -50\ K$, produced mass flow rates of $NO$ and $NO_2$ were increased. Also with decreasing temperature variation ranging from $-100\ K \leq \Delta T_{cool} \leq -50\ K$, power output was increased.

**Commented [WU1]:** اینجام همینطور .

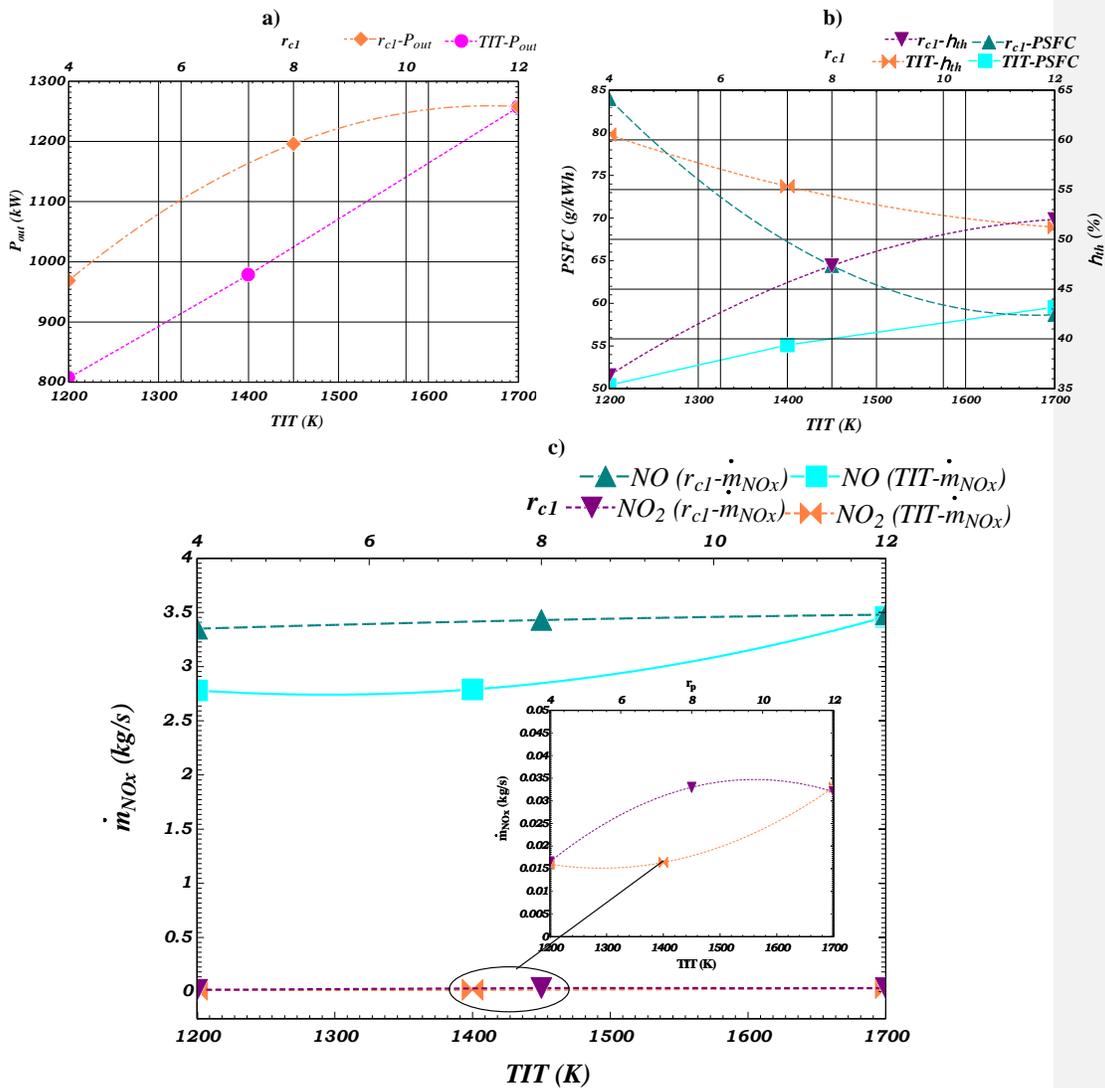

**Fig. 4.** Effect of the TIT on a) Power output b) PSFC and Thermal Efficiency c) NOx emission mass flow rate

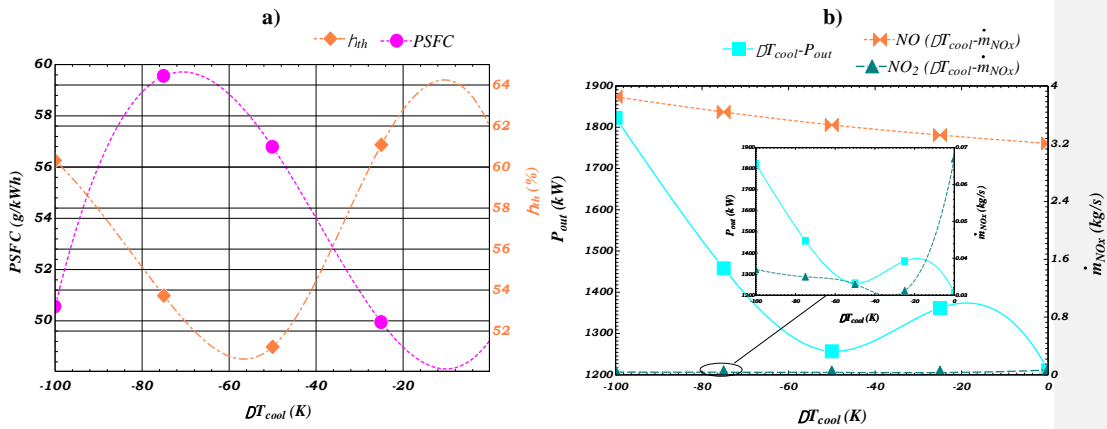

**Fig. 5.** Impacts of inlet air temperature variation on a) PSFC and Thermal Efficiency b) Power output and NOx emission mass flow rate

The output power variations with accessory cycle compression ratio ($CR$) and turbine 2 temperature ratio ($TF_{T2}$) are shown in Fig. 6a. The results were exposed an increase in $rc_1$, the output power was increased. Also, the results showed that with an increase in $TF_{T2}$, the output power is decreased.

Correspondingly, the changes in power-specific fuel consumption and thermal efficiency with rc1 and $TF_{T2}$ are shown in Fig. 6b. and the results evidenced an increase in $CR$, thermal efficiency was increased and $PSFC$ is decreased; also, by increasing $TF_{T2}$, the thermal efficiency was decreased and $PSFC$ is increased.

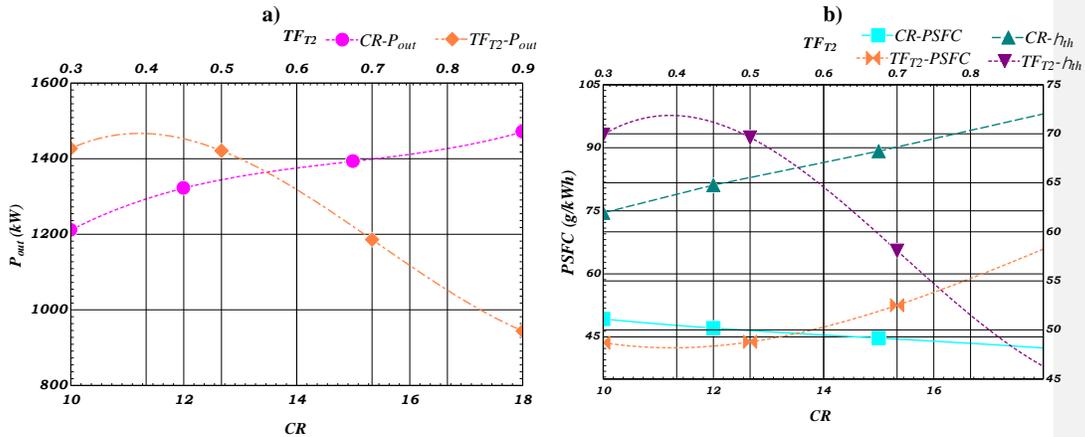

**Fig. 6.** Effect of $TF_{T2}$ and $CR$ on a) Power output b) PSFC and Thermal Efficiency

According to the recent consequences based on the thermodynamical and environmental results of the Novel Regenerative turboshaft engine combine cycle with adding inlet air cooling, the affection of Flight Altitude, Flight Mach Number, TIT, difference inlet air temperature, temperature Fraction of turbine 2, and pressure ratio of main cycle compressor has been considered. These seven parameters are provided as the input variables of the DNN in the input layer. Also, the thermal efficiency and NOx emission mass flow rate are provided as the output variables of the DDN in the output layer. In this regard, 4899 samples have been given for the first analysis. So, 20% of the dataset was set for the training of the model; so, the other data was applied for the testing of the model.

The dependency of input and output parameters are illustrated in Fig. 7 just for watching the dynamic behavior of the system to choose experiencedly the hidden layers and its neurons.

The dataset was provided in EES and DDN was modeled in the environment of Python 3.9 as an open-source coding system. First, the datasets have been normalized to the order of zero to one. Then, Adaptive moment estimation (ADAM) was chosen as the optimizer of the model and the MSE was chosen as the loss function. Also, the hidden layer activated with the rectified linear unit (Relu) function that rectified the neurons for training in the hidden layer. However, the activation function must be chosen differently at the output layer to consider both negative and positive values. The common activation function for the output layer is the sigmoid function.

Relu function can be determined as;

$$f(x) = x^+ = max(0, x) = \begin{cases} 0 & for \quad x \leq 0 \\ x & for \quad x > 0 \end{cases} \tag{55}$$

Also, the Sigmoid function can be considered as;

$$f(x) = \frac{1}{1 + e^{-x}} \tag{56}$$

The plot of these activation functions is illustrated in Fig. 8.

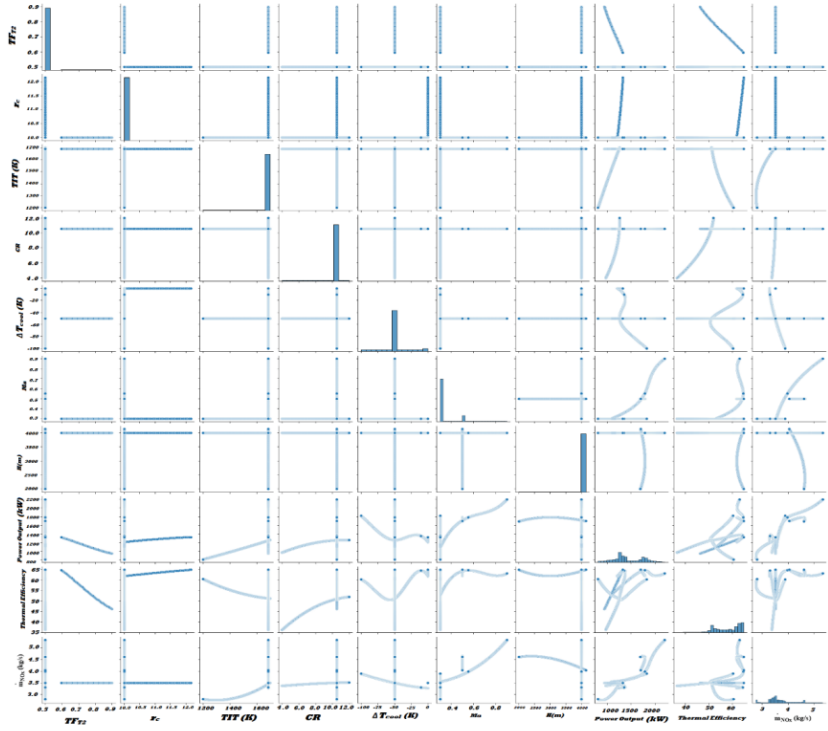

**Fig. 7.** The associations of the input and output Parameters

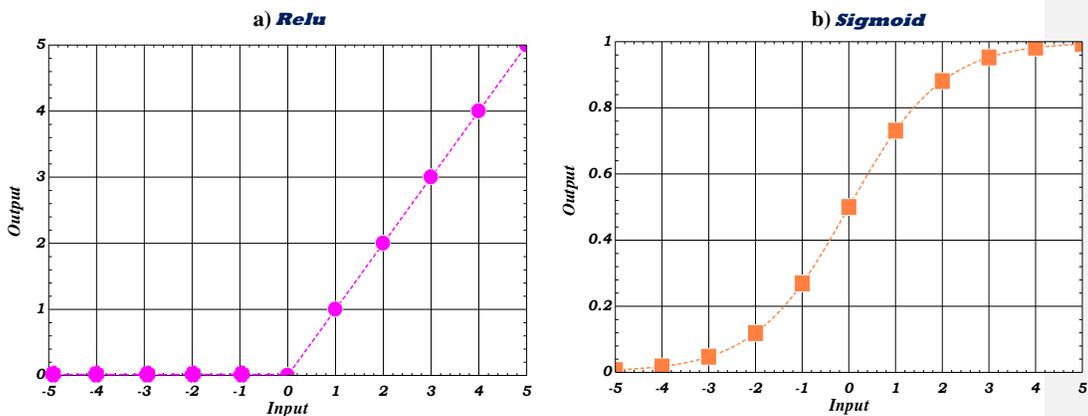

**Fig. 8.** a) Relu activation Function b) Sigmoid activation function

As previously discussed, MSE is considered the loss function. In Fig. 9, the loss function and validation lost versus epochs are presented. The results showed that at the final epochs the loss function was near zero, which allows the network to continue the process to gain the predicted values.

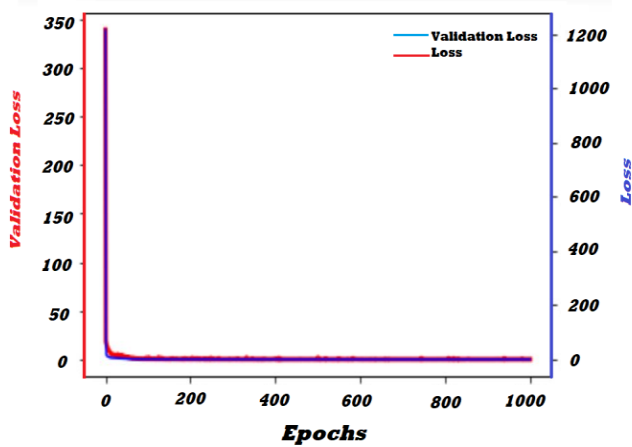

**Fig. 9.** Loss function and validation loss based on Epochs variation

In order to assess the accuracy of the integrated DNN model, the loss and cost functions such as MSE, MAE, RMSD, $R$, and $R^2$ are considered. The result showed that the model has been designed accurately with 2 hidden layers and 625 neurons for each hidden layer with 7 input parameters to predict two output parameters (Thermal efficiency and NOx emission mass flow rate). The consequences are presented in Table 2. It is tangible that the $R$ and $R^2$ are very close to 1 for both thermal efficiency and NOx emission mass flow rate for both validations of thermal efficiency and NOx emission mass flow rate prediction values with its training and its testing data.

The results reported that $R$ value for thermal efficiency prediction was calculated at 0.99973 and 0.99969 for validation training and testing datasets, respectively. Also, it has been calculated 0.99827 and 0.99806 for validating NOx emission mass flow rate prediction with training and testing datasets, respectively. $R^2$ value for thermal efficiency prediction was calculated at 0.99664 and 0.99683 for validation training and testing datasets, respectively. Also, it has been calculated 0.99652 and 0.99611 for validating NOx emission mass flow rate prediction with training and testing datasets, respectively.

Also, MSE, MAE, and RMSD of thermal efficiency demonstrated that predicted data of thermal efficiency has good agreement with the real values. Mutually, the amounts of these cost functions are calculated for both predicted parameters with validation of training and testing datasets. For instance, MSE, MAE, and RMSD of thermal efficiency for validation of training data sets were calculated at 0.13193, 0.33280, and 0.36322, respectively. Also, these parameters were calculated for thermal efficiency validation of testing data: 0.12763, 0.32442, and 0.35725, respectively.

Also, the loss functions such as MSE, MAE, and RMSD are very near to Zero for the Prediction of NOx Emission mass flow rate which proves the high accurate power of the prediction model for both testing and training data. Accordingly, MSE, MAE, and RMSD of NOx Emission mass flow rate for validation of training data sets were premeditated at 0.00105, 0.01724, and 0.03244, respectively. Also, these parameters are intended for NOx Emission mass flow rate validation of testing data: 0.00114, 0.01742, and 0.03372, respectively.

**Table 2.** matched factor of predicted parameters

|  | Thermal Efficiency | | NOx Emission mass flow rate | |
|---|---|---|---|---|
|  | Train | Test | Train | Test |
| MSE | 0.13193 | 0.12763 | 0.00105 | 0.00114 |
| MAE | 0.33280 | 0.32442 | 0.01724 | 0.01742 |
| RMSD | 0.36322 | 0.35725 | 0.03244 | 0.03372 |
| $R$ | 0.99973 | 0.99969 | 0.99827 | 0.99806 |
| $R^2$ | 0.99664 | 0.99683 | 0.99652 | 0.99611 |

The prediction model was evaluated with the training set and testing set for both output parameters of DNN. In Fig. 10, the prediction values are compared with the real data from the test and train datasets. The real values and predicted values are set at the midline of $Real = Predicted$, which shows the target predicated values are matched to the real values that gained from the system analysis. Eventually, the results showed that the prediction was successfully evaluated for this concept.

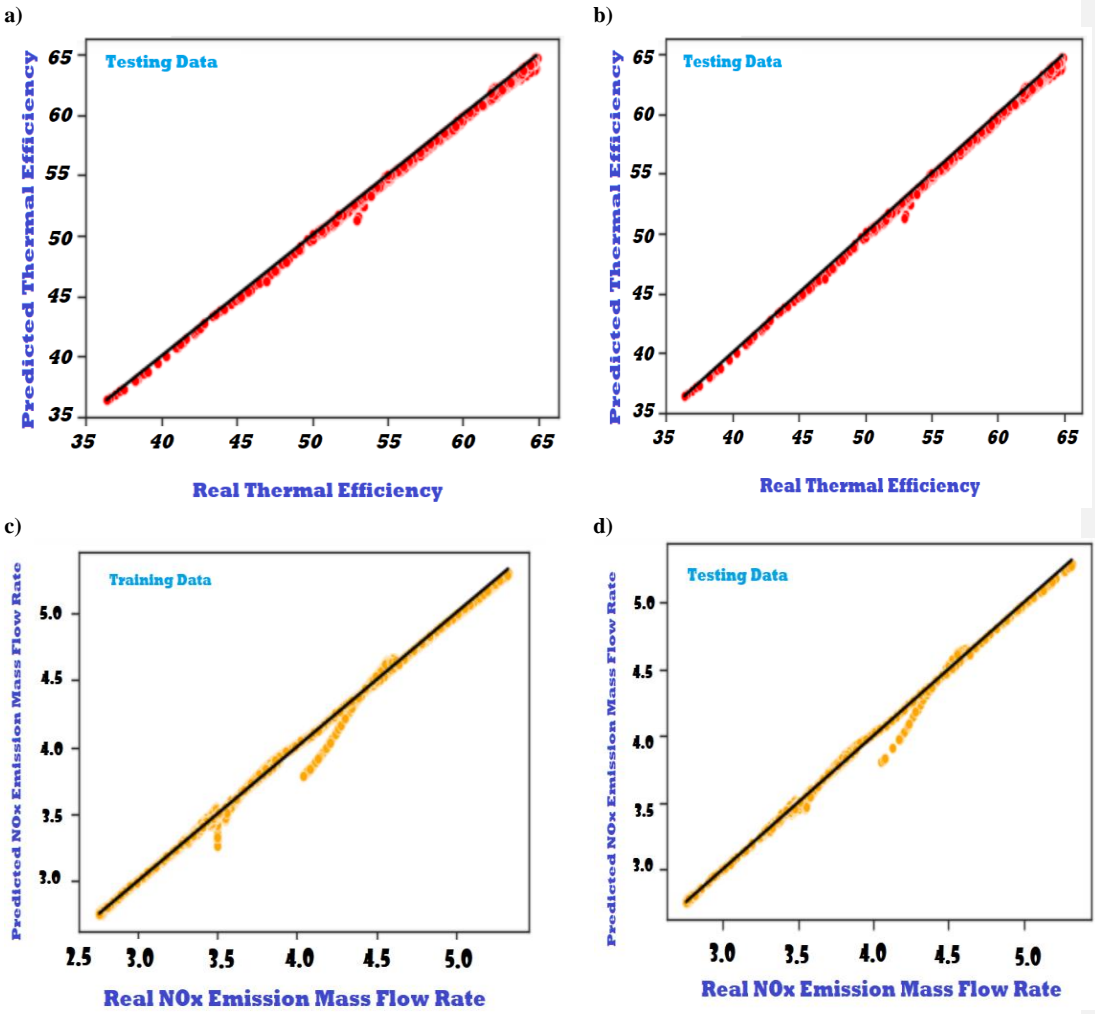

Fig. 10. Predicted output values against Real output values

## 4. Conclusion

In this integrated study, the effect of Flight-Mach number, flight altitude, and the difference between inlet air temperature of the compressor with the ambient temperature on operating parameters of the novel Regenerative turboshaft combine cycle equipped with inlet air cooling containing, power output, thermal efficiency, and Nitride oxide emission mass flow rate including produced mass flow rate of $NO$ and $NO_2$ with the use of hydrogen as fuel investigated. Also, the $NO$ and $NO_2$ produced mass flow rate, power-specific fuel consumption, and thermal efficiency with respect to the impact of turbine inlet temperature and the use of hydrogen as fuel investigated for a novel regenerative turboshaft engine equipped with inlet air cooling. Eventually, the Deep neural network was designated to predict Nitride oxide emission mass flow rate and thermal efficiency.

The most significant achievements are expressed as follows:

1. with cooling inlet air, in the range of $\Delta T_{cool} =- 50°C$ to $\Delta T_{cool} = -100°C$, specific-power fuel consumption was decreased and thermal efficiency was increased. Also, with inlet air cooling, in the range of $\Delta T_{cool} = - 25°C$ to $\Delta T_{cool} = -100°C$, $NO$, and $NO_2$ the produced mass flow rate was increased.
2. The results showed that with the increase in the Turbine 1 inlet temperature, the output power and power-specific fuel consumption were increased, but the thermal efficiency was reduced.
3. Also, by increasing the compression ratio of the main cycle compressor, the output power, and the thermal efficiency were increased but power-specific fuel consumption was decreased.
4. Also, with increasing turbine inlet temperature of the main cycle and the compression ratio of the main cycle (compressor compression ratio), the $NO$, and $NO_2$ the produced mass flow rate was increased.

5. In deep neural network modeling, both Pearson correlation and determination factors are very close to 1 for both Thermal Efficiency and NOx Emission mass flow rate to validate its training and its testing data. So, Pearson correlation factor for thermal Efficiency prediction and NOx emission mass flow rate prediction was calculated at 0.99969 and 0.99806 to validate the testing data set, with training and testing datasets, respectively. Also, the determination factor for Thermal Efficiency prediction and NOx emission mass flow rate prediction was calculated at 0.99973 and 0.99827 to validate the testing data set, respectively.

6. Mutually, the amount of mean-squared error, mean-absolute error, and root mean squared deviation for the testing and training data set was approved as the proper exact of deep neural network prediction. The amount of these parameters was calculated for validation of thermal efficiency testing data at 0.12763, 0.32442, and 0.35725, respectively. Also, it was calculated for validation of NOx emission mass flow rate testing data: 0.00114, 0.01742, and 0.03372, respectively.

**Declaration of Competing Interest**

The authors declare that they have no known competing financial interests or personal relationships that could have appeared to influence the work reported in this paper.